\documentclass[proof]{pasj01}
\Received{$\langle$reception date$\rangle$}
\Accepted{$\langle$acception date$\rangle$}
\Published{$\langle$publication date$\rangle$}

\begin{document}

\title{\textcolor{black}{Lowest Earth's atmosphere layers probed during a lunar eclipse}}
\author{Kiyoe Kawauchi$^1$, Norio Narita$^{2,3,4}$, Bun'ei Sato$^1$, Teruyuki Hirano$^1$, Yui Kawashima$^{5,6}$, Taishi Nakamoto$^1$,Takuya Yamashita$^{2,3}$, and Motohide Tamura$^{2,3,4}$}%
\altaffiltext{1}{Department of Earth and Planetary Sciences, Tokyo Institute of Technology, 2-12-1 Ookayama, Meguro-ku, Tokyo 152-8551, Japan}
\altaffiltext{2}{Department of Astronomy, The University of Tokyo, 7-3-1 Hongo, Bunkyo-ku, Tokyo
113-0033, Japan}
\altaffiltext{3}{National Astronomical Observatory of Japan, 2-21-1 Osawa, Mitaka, Tokyo 181-8588, Japan}
\altaffiltext{4}{Astrobiology Center, NINS, 2-21-1 Osawa, Mitaka, Tokyo 181-8588, Japan}
\altaffiltext{5}{Department of Earth and Planetary Science, Graduate School of Science, The University of Tokyo, 7-3-1 Bunkyo-ku, Tokyo 113-0033, Japan}
\altaffiltext{6}{Earth-Life Science Institute, Tokyo Institute of Technology, 2-12-1 Ookayama, Meguro-ku, Tokyo 152-8550, Japan}
\email{kawauchi.k.ab@m.titech.ac.jp}


\maketitle

\begin{abstract}
We report the results of detailed investigation of the Earth's transmission spectra during the lunar eclipse on UT 2011 December 10. The spectra were taken by using the High Dispersion Spectrograph (HDS) mounted on the Subaru 8.2 m telescope with unprecedented resolutions both in time and wavelength (300 s exposure time in umbra and 160,000 spectral resolution, respectively). In our penumbra and umbra data, we detected the individual absorption lines of $\rm O_2$ and $\rm H_2O$ in transmission spectra and found that it became deeper as the eclipse became deeper. It indicates that the sunlight reaching the Moon passed through lower layers in the Earth's atmosphere with time because we monitored a given point on the Moon during the full eclipse duration. From the comparison between the observed and theoretically constructed transmission spectra, the lowest altitude at which the sunlight actually passed through the atmosphere is estimated to be about 10 km from the ground, which suggests the existence of sunlight blocking clouds below that altitude. Our result would be a test case for future investigations of atmospheric structure of Earth-like exoplanets via transmission spectroscopy including the refraction effect of the planetary atmosphere.
\end{abstract}

\section{Introduction}

\begin {figure} [htbp]
 \begin{center}
  \includegraphics[width=80mm] {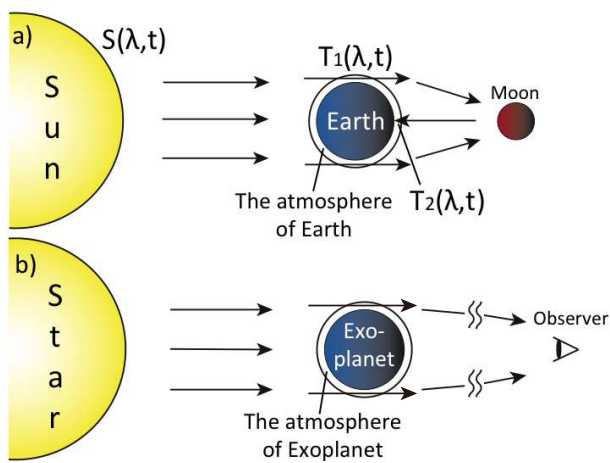}
 \end{center}
 \caption{(a) A configuration of a lunar eclipse and (b) a configuration for transmission spectroscopy of a transiting exoplanet. $S(\lambda, t)$, $T_1(\lambda,t)$, and $T_2(\lambda,t)$ are solar spectrum, telluric transmission spectrum, and telluric spectrum above a telescope, respectively.}
\label{fig:one}
\end{figure}

Transmission spectroscopy is one of the useful ways to inquire into exoplanet's atmosphere. This method searches for additional absorption features due to planetary atmospheres in host star's spectrum; one can detect some species in the planetary atmosphere by observing the stellar light through the atmosphere of planet during a transit (Figure~\ref{fig:one}). 
The detection of Na in the transmission spectrum of the hot Jupiter HD209458b was first reported by Charbonneau et al. (2002) using Hubble Space Telescope. Some other species like neutral hydrogen (e.g. Vidal-Madjar et al. 2003) and $\rm H_2O$ (e.g. Deming et al. 2013) were subsequently detected for the planet. Species such as Na and $\rm H_2O$ were also detected in other hot Jupiters like HD189733b and  HAT-P-1b (McCullough et al. 2014; Wakeford et al. 2013). 
These results are mainly observed by space telescopes, but 
ground-based observations have also played an important role in this field; detection of Na for HD209458b by Snellen et al. (2008), followed by detection of Ca and Sc for the same planet (Astudillo-Defru and Rojo 2013). 

Although the observations described above are for the hot Jupiters, it is particularly interesting to examine terrestrial exoplanets' atmospheres in order to search for biotic atmospheric species, so-called ``biomarkers'', like oxygen and ozone. For this purpose, since the Earth is currently the only planet where life exists to our knowledge, measuring Earth's transmission spectra could be the first step toward future observations of habitable terrestrial exoplanets. We can obtain a transmission spectrum of the Earth's atmosphere by observing sunlight reflected by the Moon during a lunar eclipse; the sunlight passes through the Earth's atmosphere before reaching the Moon (see Figure~\ref{fig:one} ; e.g. Pall{\'e} et al. 2009).

 By analyzing the transmission spectra of Earth-like planets, refraction of the light in the atmosphere should be properly taken into account. In the case of a planet orbiting very closely to the central star like a hot Jupiter, we can observe the stellar light passing through the entire atmosphere of the planet.
However, in the case of a distant planet like the Earth orbiting the Sun, we can only observe the stellar light that transmits above a critical altitude because of the refraction effect. For this reason, the refracted transmission spectrum becomes almost flat over a wide range of wavelength even if there exist no clouds in the atmosphere (B{\'e}tr{\'e}mieux \& Kaltenegger 2014).

The stellar light which passes through the atmosphere of the planet can actually reach an observer even before the transit begins owing to the refraction in the atmosphere (Sidis \& Sari 2010). Then the stellar light transmits through different layers of the atmosphere with time, which enables us to scan the vertical structure of the atmosphere (Garc{\'{\i}}a Mu{\~n}oz et al. 2012). Although it is not possible to apply the method to Earth-like exoplanets at this stage, future ground-based and space-based telescopes will allow us to do so (Misra, Meadows \& Crisp 2014).

From the above viewpoint, the Earth offers a unique opportunity to obtain accurate transmission spectra of the atmosphere through a lunar eclipse with high temporal (i.e. spatial) resolution using an existing large telescope. The spectra help us to understand vertical structure of Earth-like exoplanets in future, and our observations become test cases to investigate dependence of transmission spectra on elevation in the atmosphere in a more general sense.

Observations of a lunar eclipse were first pioneered by Pall{\'e} et al. (2009) for optical to near-infrared wavelength range (3600-24000~$\rm \AA$) with low-resolution spectroscopy using LIRIS and ALFOSC spectrographs attached to the 4.2 m William Herschel Telescope (WHT) and the 2.56 m Nordical Optical Telescope (NOT), respectively. 
This observation resulted in a mainly qualitative identification of the Earth's atmospheric compounds, though it could not allow us to identify absorbing species in detail because of the low spectral resolution (R$\sim$920-960). 
Vidal-Madjar et al. (2010) observed the same 2008 lunar eclipse with the SOPHIE  high-resolution spectrograph (R$\sim$75,000) at the 1.93 m telescope of the OHP (Observatoire de Haute Provence) observatory covering a wavelength range of 3900-7000~$\rm \AA$. They detected broadband signatures of $\rm O_3$ and Rayleigh scattering as well as narrowband features of Na I and $\rm O_2$ at high spectral resolution. However they did not detect narrowband features of $\rm H_2O$.
Subsequently, high-resolution broadband observations (3200-10400~$\rm \AA$) using HARPS (R$\sim$115,000) at the 3.6 m La Silla telescope and UVES (R$\sim$120,000) at the 8.2 m very large telescope (VLT) were conducted in 2010 by Arnold et al. (2014), and they detected broadband absorption lines of $\rm H_2O$ and $\rm O_2$.

Two groups independently reported the observations of the lunar eclipse on UT 2011 December 10, which we also observed using Subaru telescope too. One group, Ugolnikov et al. (2013), observed it using the 1.2m telescope at the Kourovka Astronomical Observatory, Russia and fiber-fed echelle spectrograph with a wavelength range of 4100-7800~$\rm \AA$ and a spectral resolution of R$\sim$30,000. They reported the detection of spectral features of $\rm O_2$, $\rm O_3$, $\rm O_2 \cdot O_2$ Collision Induced Absorption (CIA), $\rm NO_2$ and $\rm H_2O$. The other group, Yan et al. (2015a), conducted observations with the fiber-fed echelle High Resolution Spectrograph (HRS) mounted on the 2.16 m telescope at Xinglong Station, China. The wavelength region was 4300-10000 $\rm \AA$ and a spectral resolution was R$\sim$45,000. They detected the spectral features of Rayleigh scattering, $\rm O_2$, $\rm O_3$, $\rm O_2$ $\cdot$ $\rm O_2$, $\rm NO_2$ and $\rm H_2O$, and calculated the column densities of $\rm O_3$, $\rm H_2O$ and $\rm NO_2$. They also detected the different oxygen isotopes clearly for the first time. 

Furthermore, one of the latest lunar eclipses was observed with HARPS on April 15th, 2014 (Yan et al. 2015b).  All 382 exposure frames covered  the wavelength range from 3780 to 6910 $\rm \AA$ with a spectral resolution of R$\sim $115,000. Using these lunar eclipse spectra, they demonstrated that the Rossiter-McLaughlin effect can be a useful tool to characterize exoplanet's atmosphere (McLaughlin 1924; Rossiter 1924).

We here report our independent observations of the lunar eclipse on UT 2011 December 10 using Subaru telescope and the High Dispersion Spectrograph, HDS (Noguchi et al. 2002). Based on a series of spectra with high S/N and higher temporal and spectral resolution (R$\sim$160,000) than those of the previous works, we clearly detected temporal variations in telluric transmission spectra, which are attributed to the vertical structure of the Earth's atmosphere.

The rest of the paper is organized as follows. 
In Sections 2 and 3, we describe our observations and a procedure to extract telluric transmission spectra. In Section 4, we develop theoretical transmission spectra of the Earth's atmosphere which are compared with the observed ones. In Section 5, we discuss temporal variations in the observed transmission spectra and the difference between observed and model transmission spectra. Section 6 is devoted to the summary of this work.

\section{Observations}

We observed a total lunar eclipse on UT 2011 December 10 with the Subaru 8.2 m telescope and the High Dispersion Spectrograph (HDS; Noguchi et al. 2002), which is placed at a Nasmyth focus. 
The wavelength range was set to cover 5500-8200 $\rm \AA$, which is a non-standard format of HDS, and the slit width was set to 0.2" corresponding to a spectral resolution (R = $\lambda/\Delta \lambda$) of $\sim $160,000. This resolving power is higher than those of the previous works.

\begin {figure} [htb]
 \begin{center}
  \includegraphics[width=80mm] {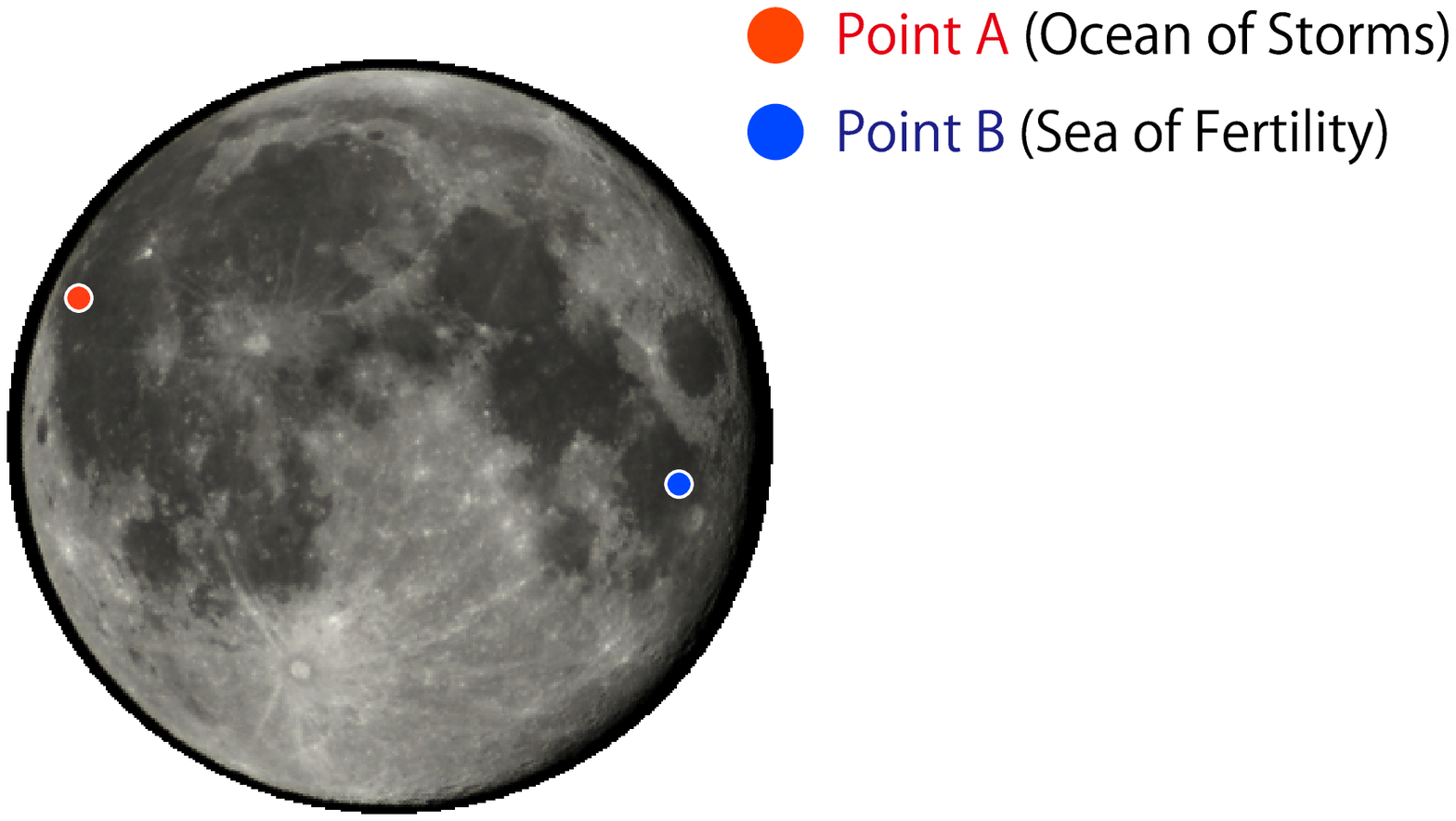}
 \end{center}
 \caption{The position of points A and B on the lunar surface.}
\label{fig:two}
 \begin{center}
  \includegraphics[width=80mm] {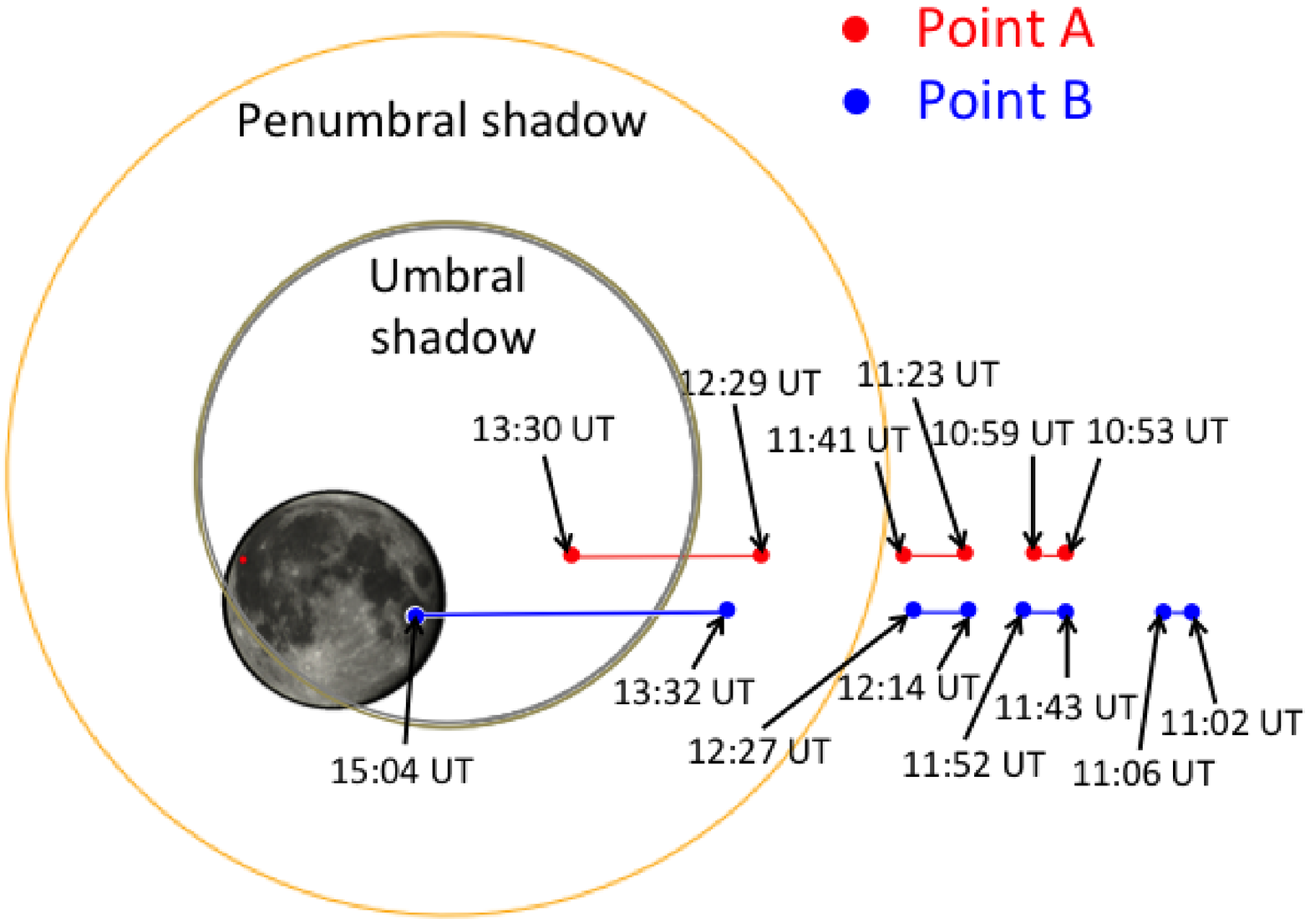}
 \end{center}
 \vspace{-8mm}
 \caption{Positional relation between Earth's shadows and the observation points on the lunar orbit from hour to hour.}
\label{fig:three}
\end{figure}

We started to observe the Moon prior to the eclipse and terminated the observation in the middle of the eclipse.
We alternately obtained a series of spectra of the sunlight reflected by two points of the lunar surface; one is Ocean of Storms (Point A) and the other is Sea of Fertility (Point B) (Figure~\ref{fig:two}). Reflectance difference of these two points is known to be small, corresponding to less hubby regions (Wilhelms \& McCauley 1971).
On point A, we took a total of 13 exposures of 30 s prior to the eclipse (i.e. full-Moon), 10 exposures of 30-300s in penumbra, and 5 exposures of 300s in umbra. On point B, we took a total of 15 exposures of 30 s prior to the eclipse (i.e. full-Moon), 6 exposures of 100-300 s in penumbra, and 11 exposures of 300 s in umbra. The sequence of the observations is shown in Figure~\ref{fig:three}.

We here note that the light reaching the observation point comes from an arc of the Earth's limb seen from the Moon.
We found that the point which the sunlight passes in the Earth's atmosphere is over the region from the center of Southern Indian ocean ($\sim $ 20$^\circ$ S, 80$^\circ$ E) to the Northern coast of Antarctica ($\sim $ 65$^\circ$ S, 100$^\circ$ E) at the time of our run. The calculation method of the longitude and latitude at each time is described in APPENDIX~1 and these results are specified in Table~\ref{table:third}.

\begin{table}[hp]
 \tbl{Latitude and longitude at the typical Universal Time during eclipse}{%
  \begin{tabular}{ccc}
  \hline
   \multicolumn{1}{c}{Universal Time} & Latitude & Longitude \\
   (UT) & ($^\circ$) & ($^\circ$)\\
   \hline
   &&\\
&   pointA &\\
\\
12:47 & 15.9 S & 81.6 E\\
12:54 & 17.2 S & 80.4 E\\
13:01 & 18.8 S & 79.7 E\\
13:07 & 20.4 S & 78.6 E\\
13:14 & 22.5 S & 77.9 E\\
13:21 & 24.9 S & 77.5 E\\
13:27 & 28.1 S & 77.5 E\\
\\
&   pointB  &\\
\\
13:48 & 28.8 S & 72.7 E\\
13:54 & 30.8 S & 72.3 E\\
14:01 & 33.5 S & 72.1 E\\
14:08 & 36.6 S & 72.4 E\\
14:14 & 39.6 S & 73.0 E\\
14:21 & 43.6 S & 74.3 E\\
14:27 & 47.4 S & 76.3 E\\
14:34 & 52.3 S & 80.0 E\\
14:41 & 57.5 S & 86.2 E\\
14:47 & 61.7 S & 93.9 E\\
14:54 & 65.6 S & 106.5 E\\
15:01 & 67.1 S & 114.6 E\\
15:07 & 65.9 S & 104.9 E\\
\hline
\end{tabular}}\label{table:third}
\end{table}

The reduction of echelle data (i.e., bias subtraction, flat-fielding, scattered-light subtraction, spectrum extraction, and wavelength calibration with Th-Ar spectra) was performed using the IRAF\footnote[1]{IRAF is distributed by the National Optical Astronomy Observatories, which is operated by the Association of Universities for Research in Astronomy, Inc. under cooperative agreement with the National Science Foundation, USA.}  software package in the standard way. The reduced spectra of full-Moon, Penumbra and Umbra have SNRs of  $\sim $ 173, $\sim $ 89 and $\sim $ 60 per pixel, respectively.
These SNR values (for point A at least) are probably biased by light scattered from the bright moon (Penumbra or full-Moon) and that it is discussed in Section 5.1.

\section{Extraction of Telluric Transmission Spectra}

A lunar-eclipse spectrum is composed of three kinds of spectra: solar spectrum $S$($\lambda, t$), telluric transmission spectrum $T_1$ ($\lambda, t$), and telluric spectrum above a telescope $T_2$ ($\lambda, t$) (Figure~\ref{fig:one}). Since we are interested in telluric transmission spectrum $T_1$($\lambda,t$), we need to eliminate the unnecessary spectra ($T_2$($\lambda,t$) and $S$($\rm \lambda,t$)) from the lunar-eclipse spectrum. 

\subsection{Eliminating $T_2$($\lambda,t$)}

\subsubsection{A Template Spectrum of $T_2$($\lambda,t$)}

We used a high-resolution spectrum of the rapid rotator HR8634 (B8V, $v \sin i =$185 kms$^{\rm -1}$; Glebocki et al. 2000), which was taken on the same night as the lunar-eclipse observations, as a template of $T_2$. 
Since the star is rapidly rotating, stellar absorption lines in the spectrum are largely broadened and the spectrum is almost featureless. Thus the spectrum basically conveys only the telluric spectrum above a telescope, and we can use it as a template of $T_2$. HR8634 is also known to have fewer interstellar absorption lines,  which makes the star an appropriate target for this purpose.
Using this obtained template spectrum, we next evaluate the dependence of $T_2$($\lambda,t$) on airmass and wavelength shift of $T_2$ in order to match the template spectrum with actual $T_2$ in lunar-eclipse spectra.

\subsubsection{Airmass Variation}

$T_2$($\lambda,t$) is dependent on airmass, $AM$ (= $\sec(z)$; $z$ is zenith distance).
Assuming the plane parallel atmosphere of the Earth, $T_2$ can be expressed as
\begin{equation}
  T_2(\lambda, t) = T_2(\lambda,t_0)^{AM/AM_0},
\end{equation}
where $T_2$($\lambda,t_0$) is the template spectrum taken at time $t_0$ and airmass $AM_0$, and $T_2$($\lambda,t$) is an expected telluric spectrum above a telescope taken at time $t$ and airmass $AM$ assuming that $T_2$ was constant for the rest of the night. Based on that relation, we can estimate line depth of $T_2$ inherent to each lunar-eclipse spectrum.
The time variation of the precipitable water vapour (PWV) can cause the depth variation in $\rm H_2O$ absorption lines. This effect, however, is negligible in this case as the value of PWV obtained from the data of Caltech Submillimeter Observatory \footnote[2]{http://cso.caltech.edu/tau/} is small and has little change during observation (PWV $\sim $ 0.64 - 1.70 mm).

 We adopted a high-S/N spectrum of HR8634 as $T_2$($\lambda,t_0$), which was obtained by averaging 8 spectra of the star taken prior to the lunar eclipse in a short period of time with negligible change in airmass $<$ 0.02. The mean time and airmass were 8:35 UT and 1.03, respectively, which were attributed to the $T_2$($\lambda,t_0$).
\subsubsection{Wavelength shift}

Variations of absorption lines in $T_2$($\lambda,t$) are found not only in line depth but also in wavelength of the lines. The wavelength variation can be caused by wind in the atmosphere, error of wavelength calibration, instability of the spectrograph, and so on. 
The Earth's rotation velocity is about 0.5 kms$^{\rm -1}$. The wavelength variation caused by the velocity (Doppler shift) is about $\rm \pm$ 0.0055 $\rm \AA$ at a wavelength of 6000 $\rm \AA$. The typical error of wavelength calibration is about 0.0011 $\rm \AA$ for HDS and there could also be temporal variation due to instrumental instability. Unfortunately, we cannot discriminate among these causes, and cannot directly derive the wavelength shift in umbra spectra since the telluric transmission spectrum $T_1$ is easily confused with $T_2$ in umbra spectra.

Alternatively, we estimated wavelength shifts of $T_2$ in umbra spectra by using those in the full-Moon spectra taken outside of the eclipse.  We measured wavelength shifts of telluric $\rm O_2$ lines between 6306.45 $\rm \AA$ and 6306.66 $\rm \AA$ in full-Moon and Penumbra spectra relative to those in the template $T_2$, and tried to extrapolate the values in umbra by using polynomials.
Figure~4 shows the derived wavelength variations of $T_2$ together with the fitted polynomials as a function of time and these results are specified in Table~\ref{table:first}. We found that all the polynomials gave reasonable estimates of the wavelength shifts in umbra for Point A, but this was not the case with Point B. Thus we adopted the values derived by  a quintic equation for Point A, and the value of -0.1 kms$^{\rm -1}$, which is the average of the 4 data closest to umbra, for Point B. These choices of the fitting functions have less impact on the subsequent analyses and results.

Figure~5 and 7 shows a resultant full-Moon and a lunar-eclipse spectrum divided by the matched $T_2$. As shown in the figure, we successfully eliminated $T_2$ from the spectra down to the photon noise level.

\begin {figure} [htb]
 \begin{center}
  \includegraphics[keepaspectratio,width=80mm] {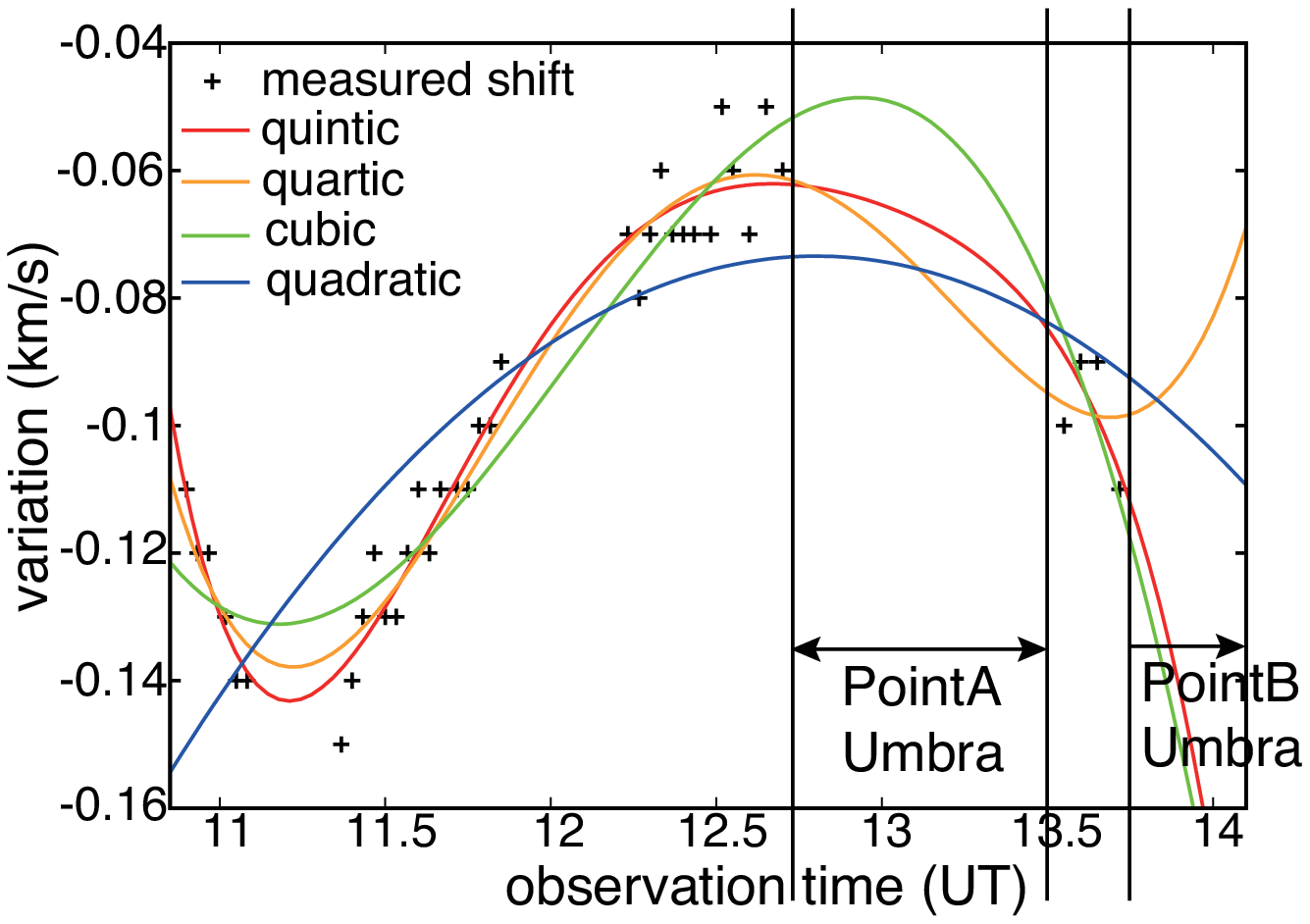}
 \end{center}
 \caption{Black pluses: temporal wavelength (velocity) shifts of $T_2$($\lambda,t$) in full-Moon and Penumbra spectra relative to the $T_2$($\lambda,t$) template spectrum. The red, orange, green, and blue lines are the best-fit quintic, quartic, cubic, and quadratic curves to the black pluses, respectively.}
\label{fig:four}
\end{figure}

\renewcommand{\arraystretch}{0.8}
\begin{table}[hp]
 \tbl{The relation between the period elapsed from the first full-Moon data and the variation in wavelength of $T_2$($\lambda,t$) from HR8634 data}{%
  \begin{tabular}{cc}
  \hline
   \multicolumn{1}{c}{Universal Time} & Variation of Wavelength \\
   (UT) & at ${\rm O_2}$ spectra (${\rm kms^{-1}}$) \\
   \hline
10:54 & -0.11   \\
10:56 & -0.12  \\
10:58 & -0.12  \\
11:01 & -0.13  \\
11:03 & -0.14  \\
11:05 & -0.14 \\
11:22 & -0.15  \\
11:24 & -0.14  \\
11:26 & -0.13  \\
11:28 & -0.12  \\
11:30 & -0.13  \\
11:32 & -0.13  \\
11:34 & -0.12  \\
11:36 & -0.11  \\
11:38 & -0.12  \\
11:40 & -0.11  \\
11:43 & -0.11  \\
11:45 & -0.11  \\
11:47 & -0.10  \\
11:49 & -0.10  \\
11:51 & -0.09  \\
12:14 & -0.07  \\
12:16 & -0.08 \\
12:18 & -0.07  \\
12:20 & -0.06  \\
12:22 & -0.07  \\
12:24 & -0.07  \\
12:26 & -0.07  \\
12:29 & -0.07  \\
12:31 & -0.05  \\
12:33 & -0.06  \\
12:36 & -0.07  \\
12:39 & -0.05  \\
12:42 & -0.06  \\
13:33 & -0.10  \\
13:36 & -0.09  \\
13:39 & -0.09  \\
13:43 & -0.11  \\
\hline
\end{tabular}}\label{table:first}
\end{table}
\renewcommand{\arraystretch}{1}

\subsection{Eliminating $S$($\lambda,t$)}

A full-Moon spectrum from which $T_2$($\lambda,t$) is eliminated is equivalent to solar spectrum $S$($\lambda,t$) assuming the moon has a perfect flat reflectance spectrum. The effect of the moon characteristic can be negligible as our observation points has little reflection characteristic.  We obtain $T_1$($\lambda,t$) by dividing the $ T_2$-eliminated lunar-eclipse spectrum by $S$($\lambda, t$).
In practice the wavelength of $S$($\lambda,t$) temporally varies according to the mutual velocities of the Sun, Earth, and Moon. Thus we correct the variations by measuring wavelength shifts of some unblended solar absorption lines in 6260.93-6261.25 $\rm \AA$, 6327.39-6327.78 $\rm \AA$, and 6470.5-6472.5 $\rm \AA$, relative to the full-Moon spectrum (Table~\ref{table:second}).

\begin{table}[hp]
 \tbl{The relation between the time elapsed from the first full-Moon data and the variation in wavelength of S($\lambda$,t) from the first full-Moon data}{%
  \begin{tabular}{ccc}
  \hline
   \multicolumn{1}{c}{Universal Time} & Wavelength shift & Wavelength shif \\
   (UT) & at ${\rm O_2}$ spectra (${\rm kms^{-1}}$) &  at ${\rm H_2O}$ spectra (${\rm kms^{-1}}$)\\
   \hline
12:29 & -0.84 &  -0.83\\
12:31 & -0.90 & -0.89\\
12:33 & -0.97 & -0.95\\
12:36 & -1.03 & -1.01\\
12:39 & -1.10 &  -1.08\\
12:42 & -1.17 &  -1.12\\
12:47 & -1.16 & -1.13\\
12:54 & -0.56 & -0.51\\
13:01 & -0.50 & -0.43\\
13:07 & -0.56 & -0.46\\
13:14 & -0.61 & -0.45\\
13:21 & -0.66 & -0.52\\
13:27 & -0.59  & -0.53\\
13:33 & -0.94 & -0.93\\
13:36 & -0.98 & -0.97\\
13:39 & -1.05 & -1.02\\
13:43 & -1.11 & -1.09\\
13:48 & -1.05 & -1.02\\
13:54 &  -0.35 & -0.35\\
14:01 & -0.18 & -0.15\\
14:08 & -0.12 & -0.07\\
14:14 & -0.09 & -0.09\\
14:21 & -0.04 & -0.05\\
14:27 & 0.00 & -0.01\\
14:34 & 0.11 & 0.10\\
14:41 & 0.12 & 0.05\\
14:47 & 0.16 & 0.30\\
14:54 & 0.51 & 0.46\\
15:01 & 0.41 & 0.62\\
15:07 & 0.57 & 0.65\\
\hline
\end{tabular}}\label{table:second}
\end{table}

Figure~6 and 8 show resultant $T_2$- and $S$-eliminated full-Moon and lunar-eclipse spectra. We successfully eliminated $S$ from the spectra well below the signal level of $T_1$.

\begin {figure} [htbp]
\begin{center}
\includegraphics[keepaspectratio,width=16cm] {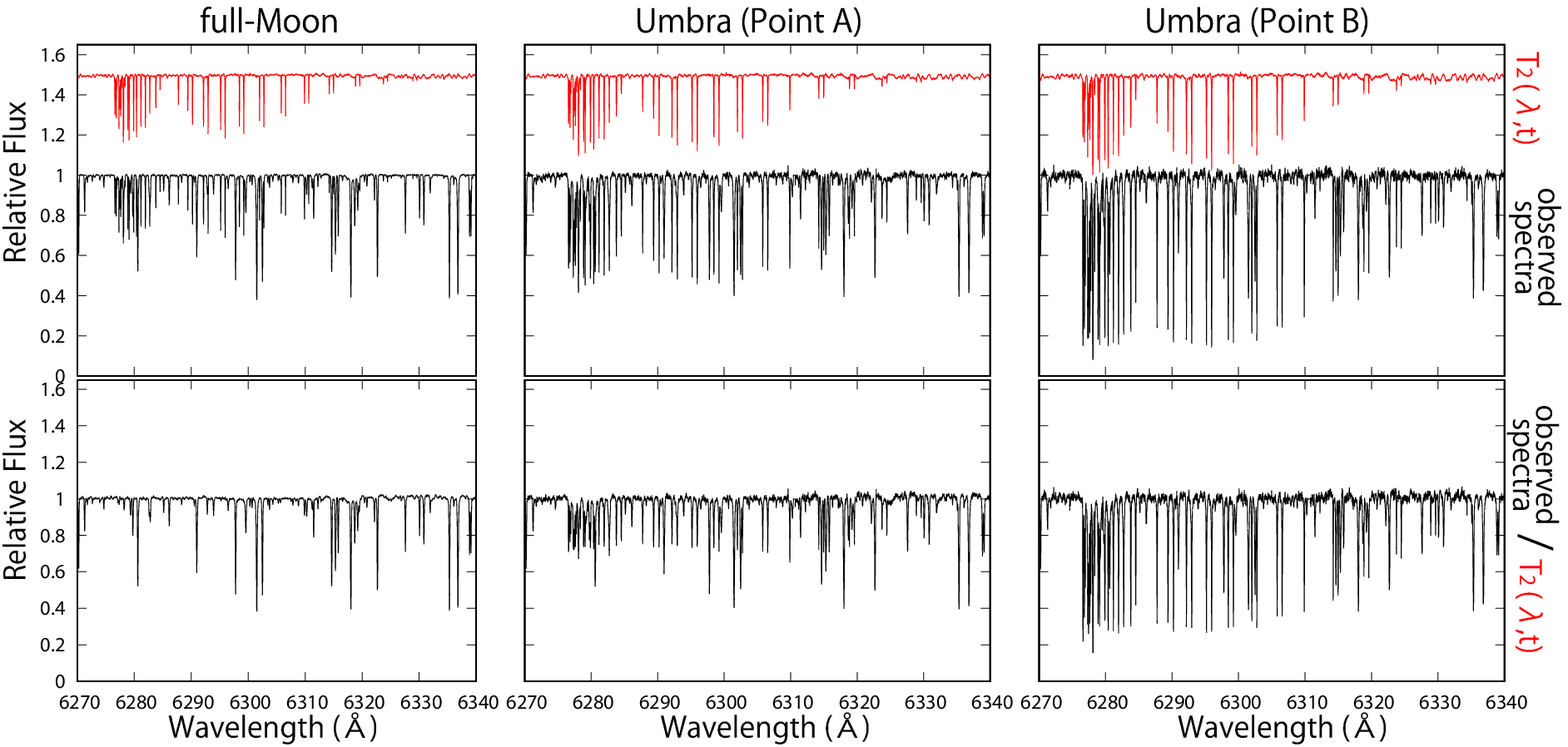}
\end{center}
\caption{{\bf Left:} a full-Moon spectrum around $\rm O_2$ lines (black line in upper panel), and the spectrum from which $T_2$($\lambda,t$) (red line in upper panel) is eliminated (bottom panel).
{\bf Center,Right:} a lunar-eclipse spectrum around $\rm O_2$ lines (black line in upper panel), and the spectrum from which $T_2$($\lambda,t$) (red line in upper panel) is eliminated (bottom panel).}
\label{fig:five}
\begin{center}
\includegraphics[keepaspectratio,width=16cm] {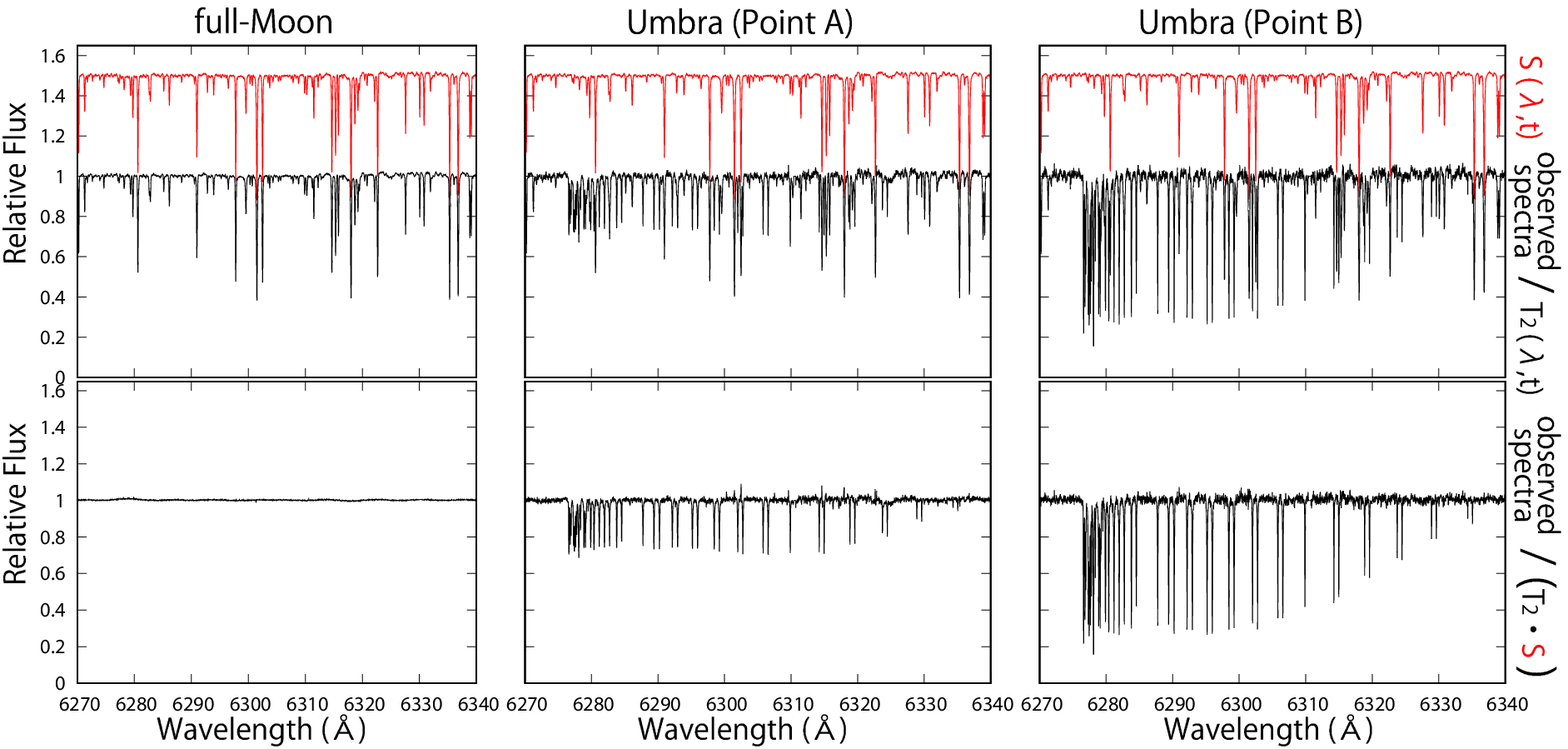}
\end{center}
\caption{{\bf Left:} a full-Moon spectrum around $\rm O_2$ lines (black line in upper panel) from which $T_2$($\lambda,t$) is eliminated, and the spectrum from which $S$($\lambda,t$) (red line in upper panel) is eliminated (bottom panel).
{\bf Center,Right:} a lunar-eclipse spectrum around $\rm O_2$ lines (black line in upper panel) from which $T_2$($\lambda,t$) is eliminated, and the spectrum from which $S$($\lambda,0$) (red line in upper panel) is eliminated (bottom panel). A telluric transmission spectrum, $T_1$($\lambda,t$), is clearly seen in the bottom panel.}
\label{fig:six}
\end{figure}
\begin {figure} [htbp]
\begin{center}
\includegraphics[keepaspectratio,width=16cm] {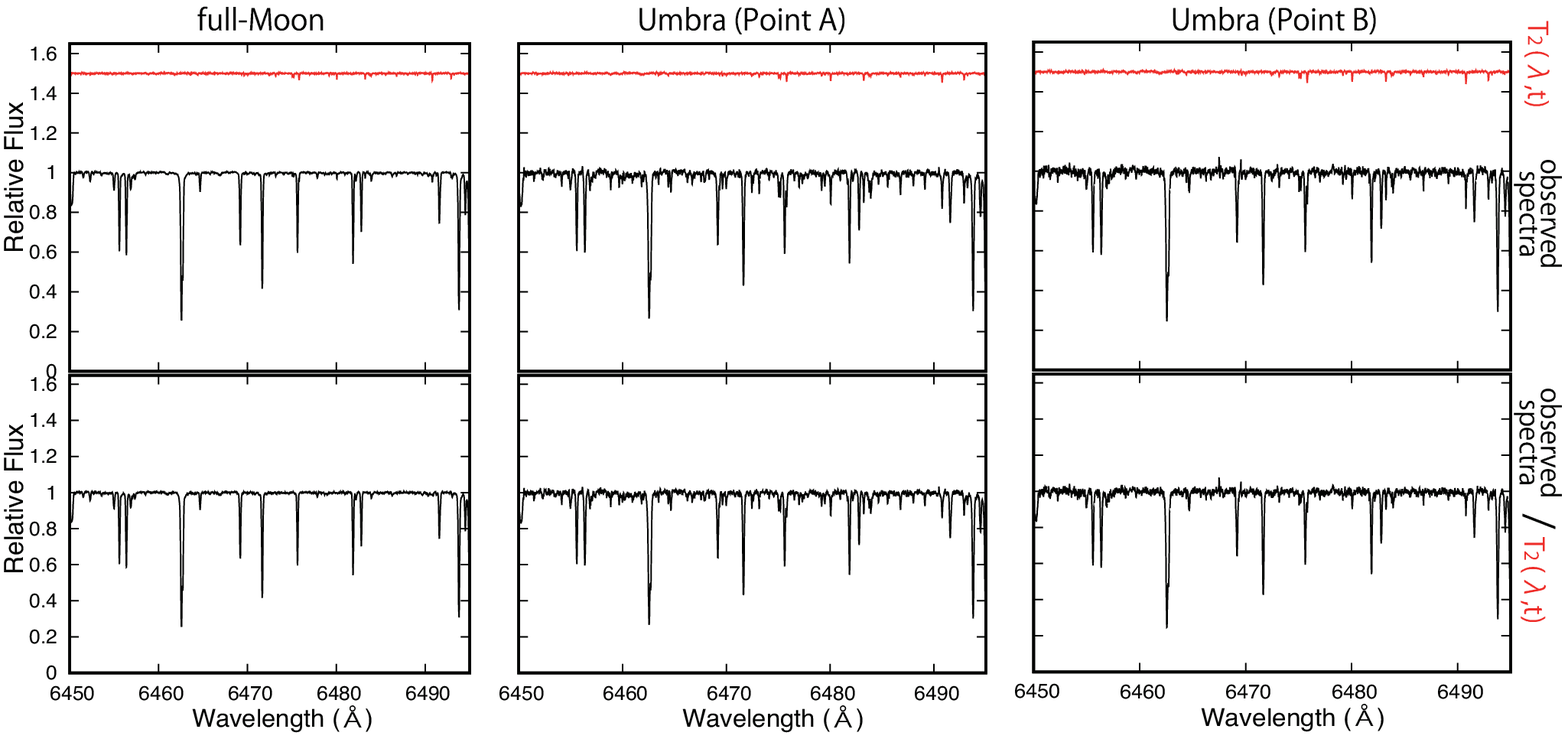}
\end{center}
\caption{{\bf Left:} a full-Moon spectrum around $\rm H_2O$ lines (black line in upper panel), and the spectrum from which $T_2$($\lambda,t$) (red line in upper panel) is eliminated (bottom panel).
{\bf Center,Right:} a lunar-eclipse spectrum around $\rm H_2O$ lines (black line in upper panel), and the spectrum from which $T_2$($\lambda,t$) (red line in upper panel) is eliminated (bottom panel).}
\label{fig:seven_1}
 \begin{center}
\includegraphics[keepaspectratio,width=16cm] {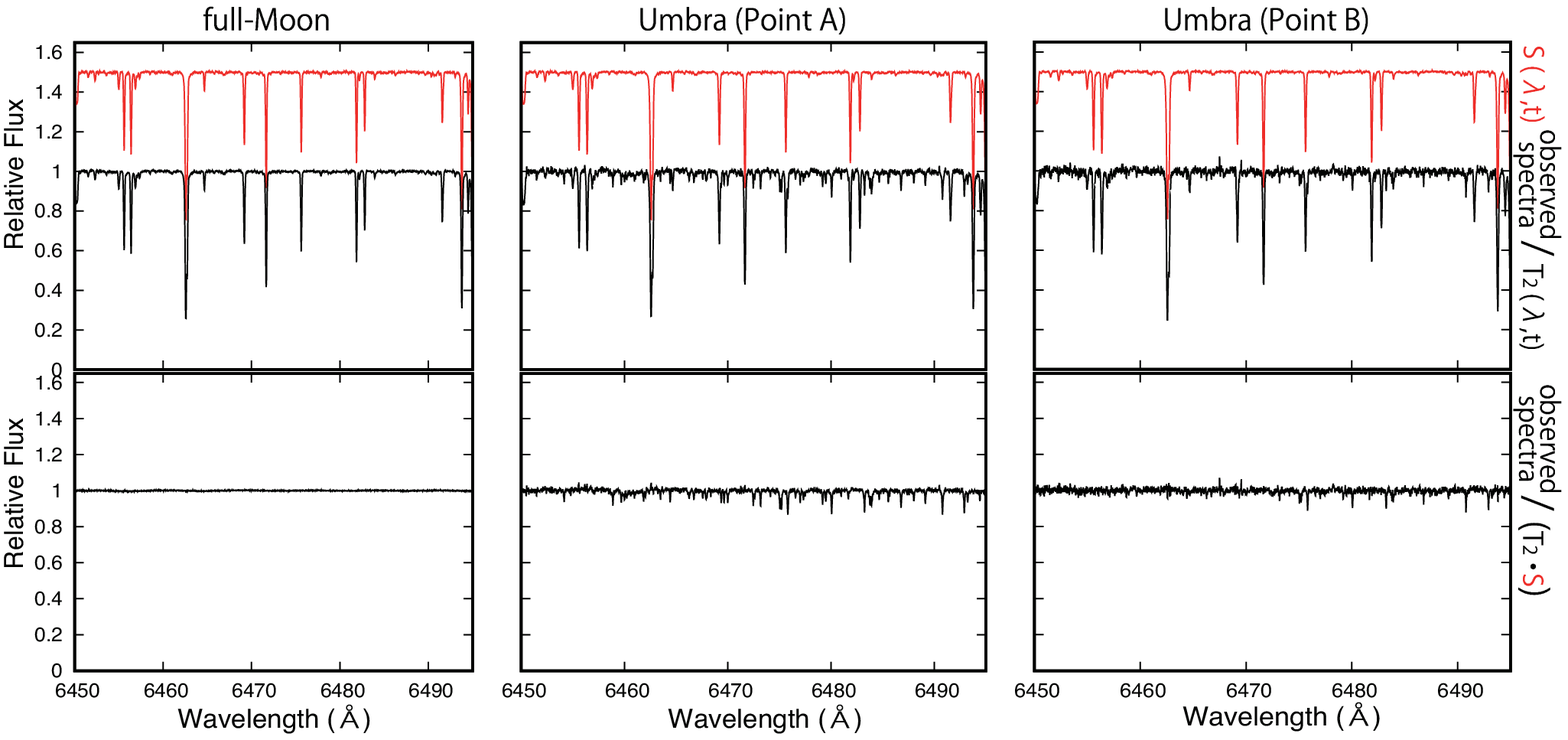}
 \end{center}
 \caption{{\bf Left:} a full-Moon spectrum around $\rm H_2O$ lines (black line in upper panel), and the spectrum from which $S$($\lambda,t$) (red line in upper panel) is eliminated (bottom panel).
{\bf Center,Right:} a lunar-eclipse spectrum around $\rm H_2O$ lines (black line in upper panel), and the spectrum from which $S$($\lambda,0$) (red line in upper panel) is eliminated (bottom panel). A telluric transmission spectrum, $T_1$($\lambda,t$), is clearly seen in the bottom panel.
}
\label{fig:seven}
\end{figure}

\subsection{$\rm O_2$ and $\rm H_2O$ transmission spectrum}

Figure~9 and 10 show thus extracted $T_1$ spectra of penumbra, and umbra. We can clearly see $\rm O_2$ absorption lines in 6275-6330 $\rm \AA$, and $\rm H_2O$ absorption lines in 5870-5930 $\rm \AA$ and 6465-6495 $\rm \AA$. We can also see  that the absorption lines become deeper as the eclipse becomes deeper.

\begin {figure} [htbp]
\begin{center}
\vspace{0mm}
\resizebox{16.0cm}{16.0cm}{\includegraphics{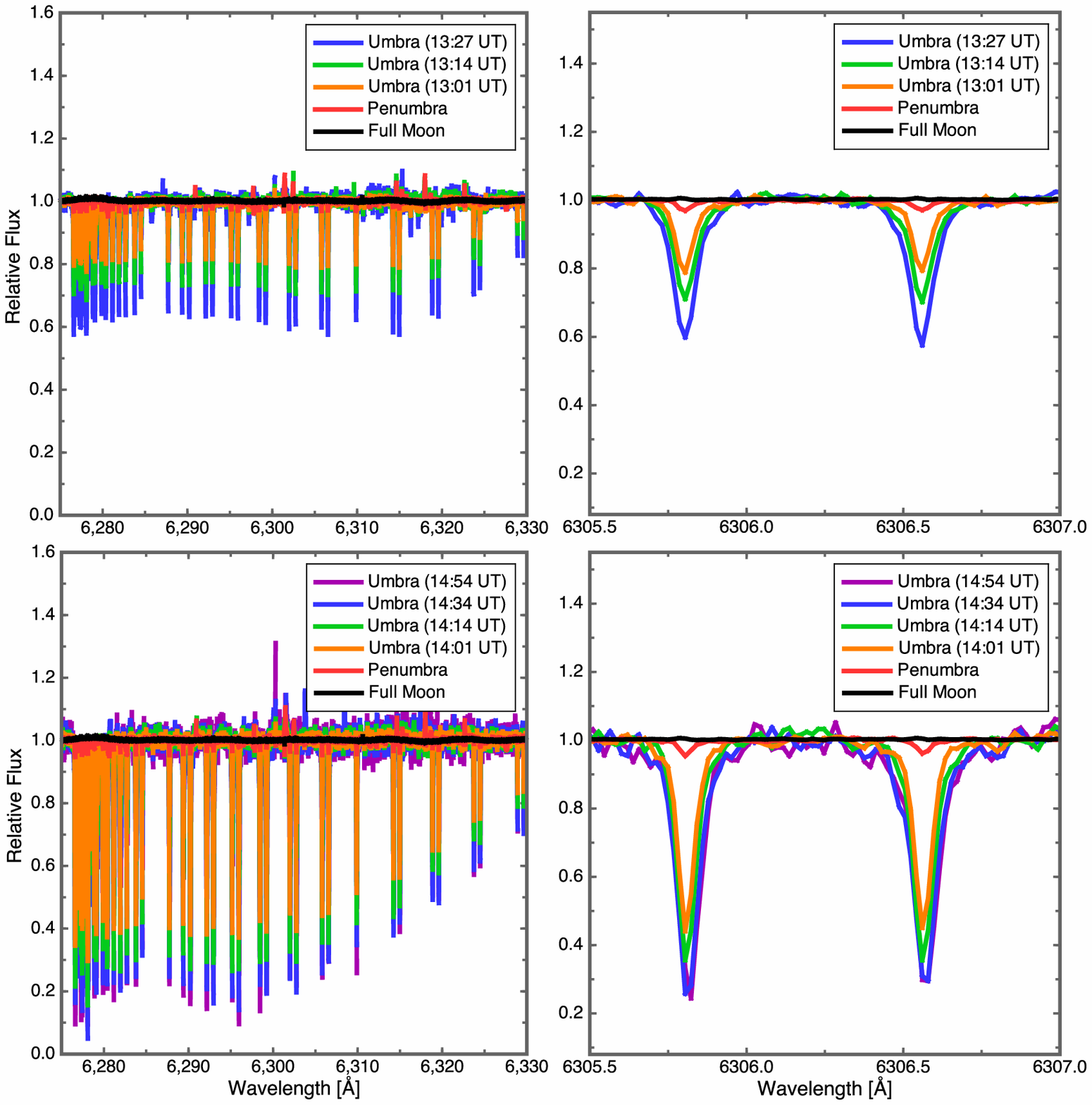}}
\end{center}
\caption{Transmission spectrum ($T_1$($\lambda, t$)) of $\rm O_2$ (left). $\rm O_2$ absorption lines of 6305.80 $\rm \AA$ and 6306.56 $\rm \AA$ are zoomed-in on the right panels. Upper figures show the spectra observed at Point A and lower figures show the spectra observed at Point B.}
\label{fig:eight}
\end{figure}
\begin {figure} [htbp]
\begin{center}
\vspace{0mm}
\resizebox{16.0cm}{16.0cm}{\includegraphics{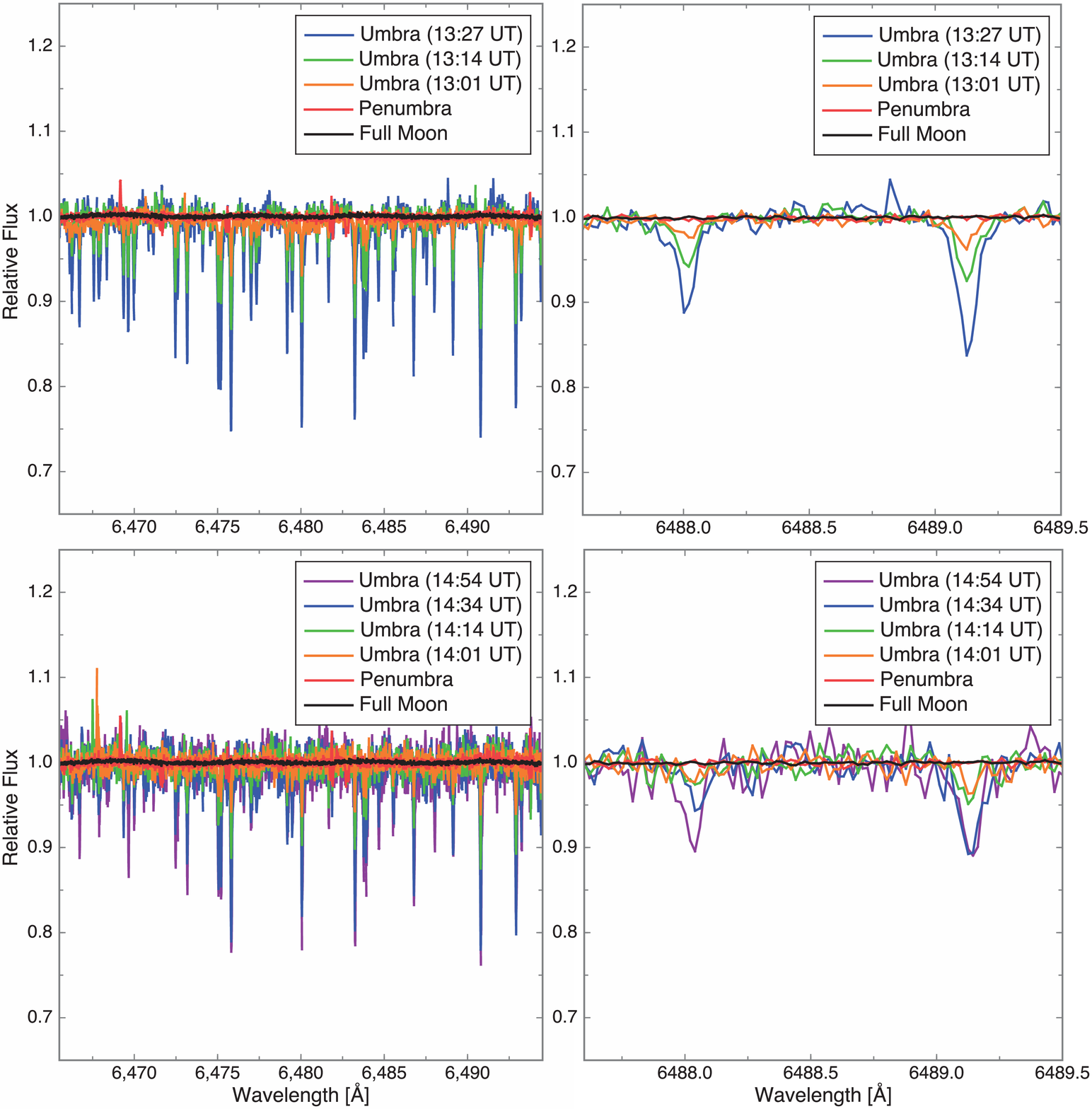}}
\end{center}
\caption{Transmission spectra ($T_1$($\lambda, t$)) of $\rm H_2O$ (left). $\rm H_2O$ absorption lines of 6487 $\rm \AA$ and 6489.5 $\rm \AA$ are zoomed-in on the right panels. Upper figures show the spectra observed at Point A and lower figures show the spectra observed at Point B.}
\label{fig:nine}
\end{figure}

\section{Model Spectra}

We found that the telluric transmission spectra vary in depth with time.
This is because we see the sunlight passing through a lower-altitude region in the Earth's atmosphere as the eclipse becomes deeper. In order to evaluate the variations quantitatively, we construct theoretical transmission spectra of the Earth and compare them with the observed ones.

\subsection{Altitude}

The lowest surface altitude of the Earth's atmosphere, $z_{min}$, through which the sunlight passes towards the Moon can be calculated from the positional relation between the Sun, Earth, and observation point on the lunar surface by using the Geometric Pinhole Model (Vollmer \& Gedzelman 2008). 

However, the model only works for umbra (not penumbra), and assumes that the refraction angle in the Earth's atmosphere is 70 arcmin for $z_{min}$ = 0 km.
The value is based on analytical approximation, but in fact it is not valid in the case of such a small $z_{min}$ (i.e. large refraction angle).
Alternatively we calculate $z_{min}$ for both umbra and penumbra with use of the 4th order Runge Kutta method following Tsumura et al. (2014).

\begin{figure}[htb]
\begin{center}
\vspace{0mm}
 \includegraphics[keepaspectratio,width=80mm] {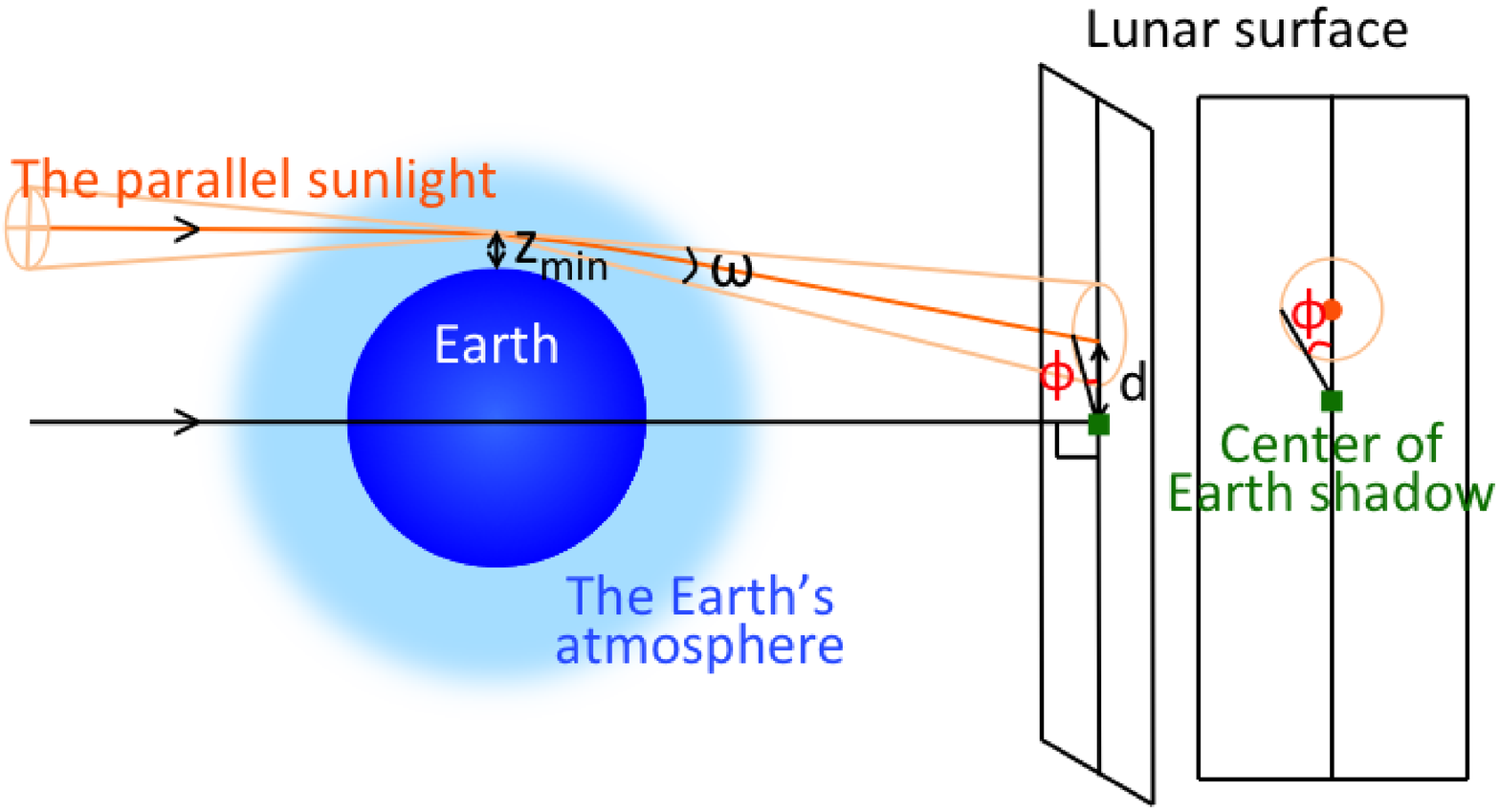}
\includegraphics[keepaspectratio,width=80mm] {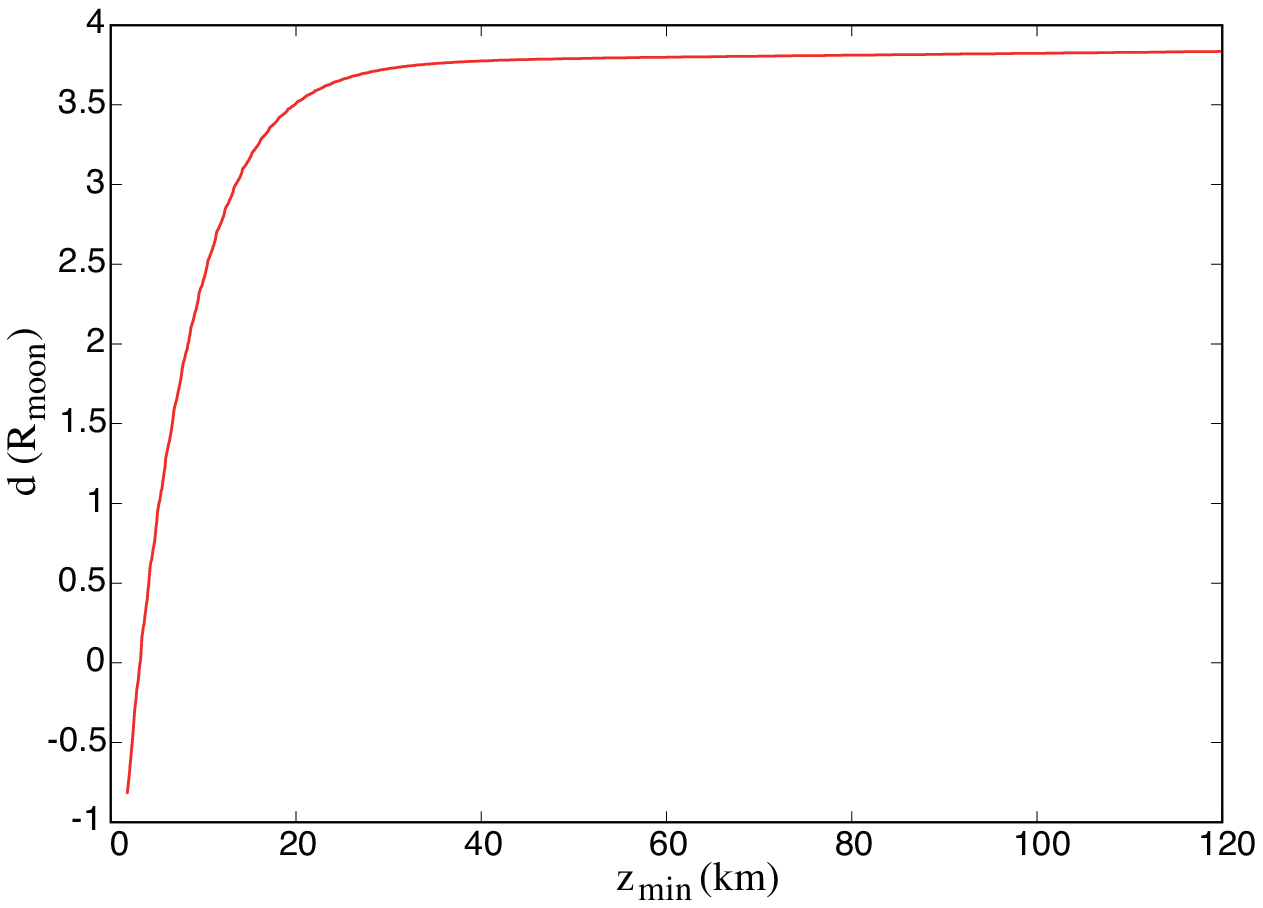}

\end{center}
\caption{Relation between the distance from the center of the Earth shadow to the point which the sunlight reaches ($d$) and the minimum altitude at which the sunlight passes in the Earth's atmosphere ($z_{min}$). $R_{moon}$ means the lunar radius.}
\label{fig:ten}
\end{figure}

First of all, we calculate and trace refraction of the sunlight that enters into the Earth's atmosphere in parallel to the line connecting the center of Earth shadow on the lunar surface with the center of the Earth (Figure~\ref{fig:ten}).
Refractive index at each point in the atmosphere is calculated from geometric optics, and rays with a variety of incident altitudes (i.e. impact parameter) are traced. The calculation is based on the assumption that the Earth's atmosphere is spherically symmetric. The number-density distribution in vertical direction is given by the T-P profile for mid-latitude in MIPAS Model Atmosphere of the Earth (2001)\footnote[3]{http://eodg.atm.ox.ac.uk/RFM/atm/} because we observed the sunlight passing over the center of Southern Indian ocean and the Northern coast of Antarctica this time.
As a result of the calculation, we obtain a distance ($d$) from the center of the Earth shadow to the point that the refracted sunlight reaches the lunar orbit. The distance, $d$, is related to the lowest surface altitude ($z_{min}$) on the Earth at which the sunlight passes. The relation between $d$ and $z_{min}$ is shown in Figure~\ref{fig:ten}. Since we know the distance from the center of the Earth shadow to the point A and B on the lunar surface from hour to hour (Appendix~1), we can determine $z_{min}$ at each observing time.

However, as the sun has a finite size, the sunlight that passes the minimum altitude actually illuminates a finite area, $\omega $, on the lunar surface (Figure~\ref{fig:ten}). The area can be considered to be circular because of the projection of the Sun. In this case, we can estimate the range of $d$ whose corresponding $\omega $ include the observation point with various position angle $\phi $ (Figure~\ref{fig:ten}) using the same method as the pinhole model. The minimum and the maximum $d$ is calculated as,
\begin{equation}
d_{max,min}(\phi ) = d_0 \cdot \cos \phi \pm \sqrt{r^2-(d_0 \sin \phi)^2},
 \end{equation}
where $d_0$ is the distance between the center of the Earth shadow and the observation point, and $r$ is the radius of the Earth shadow on the lunar surface. The $d_{max}$ and $d_{min}$ are converted into the maximum and minimum of $z_{min}$, respectively, by using the relation presented in the right panel of Figure~\ref{fig:ten}.

\begin{figure}[htb]
\begin{center}
\vspace{0mm}
\includegraphics[keepaspectratio,width=16cm]{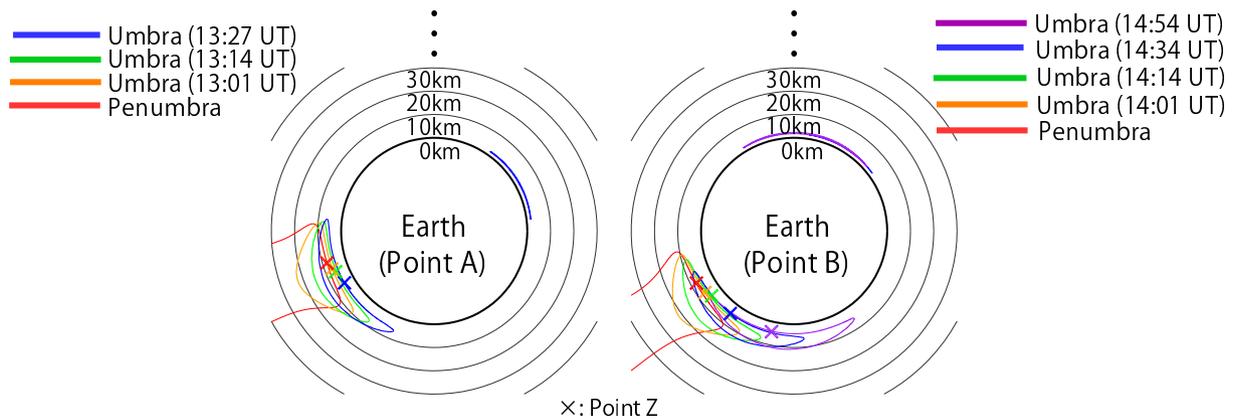}
\end{center}
\caption{The solar images as observed from the Moon at different times during lunar eclipse on point A (left) and point B (right). These images mean the ranges of the altitudes, $z_{min}$, at which the sunlight pass in the Earth's atmosphere. Point Z is the lowest point of each images.}
\label{fig:eleven}
\end{figure}

Figure~\ref{fig:eleven} shows the derived ranges of $z_{min}$ for point A (left) and point B (right) at each observing time. In this figure, we can see that (1) the sunlight reaching the observation point is transmitted through lower layers of the Earth's atmosphere with time, (2) the sunlight that is transmitted through the opposite side with respective to the Earth center also reaches the observation point at the middle of the eclipse, and (3) the bottom of the range of $z_{min}$ is not 0 km but 1.8 km (Table~\ref{table:fourth}) due to refraction.

\begin{table}[hp]
 \tbl{Altitude range at the typical Universal Time during eclipse}{%
  \begin{tabular}{ccl}
  \hline
   \multicolumn{1}{c}{observation point} & observation time & altitude range \\
   A or B & (UT) & (km)\\
   \hline
A & 12:47 & 8.1 - (penumbra)\\
A & 12:54 & 7.3 - 82.4\\
A & 13:01 & 6.6 - 22.4\\
A & 13:07 & 6.0 - 18.0\\
A & 13:14 & 5.5 - 15.2\\
A & 13:21 & 5.0 - 13.6\\
A & 13:27 & 4.5 - 12.2\\
 & & 1.8 - 2.0 (other side)\\
B & 13:48 & 7.7 - (penumbra)\\
B & 13:54 & 7.1 - 35.4\\
B & 14:01 & 6.5 - 21.6\\
B & 14:08 & 5.9 - 17.7\\
B & 14:14 & 5.6 - 15.7\\
B & 14:21 & 5.1 - 14.1\\
B & 14:27 & 4.9 - 13.2\\
B & 14:34 & 4.6 - 12.3\\
 & & 1.8 -2.0 (other side)\\
B & 14:41 & 4.3 - 11.5\\
 & & 1.8 -2.2 (other side)\\
B & 14:47 & 4.2 - 11.2\\
 & & 1.8 - 2.3 (other side)\\
B & 14:54 & 4.1 - 11.0\\
 & & 1.8 - 2.3 (other side)\\
B & 15:01 & 4.1 - 10.9\\
 & & 1.8 - 2.4 (other side)\\
 B & 15:07 & 4.1 - 11.0\\
 & & 1.8 - 2.3 (other side)\\
\hline
\end{tabular}}\label{table:fourth}
\end{table}

We can also see that the sunlight passed through the similar layers of the atmosphere for point A and B. However, the line depth of the transmission spectrum for point A is much shallower than that for point B (Figure~\ref{fig:eight}). We discuss this difference in section 5.1.

\subsection{Creating modeled spectrum}

A theoretical transmission spectrum ($T_1$($\lambda,t$)) is obtained from the Beer-Lambert law:
\begin{equation}
 T_{1} ( \lambda,t) = e^{-\int ( u \cdot \sigma )dx},
\end{equation}
where $u$, $\sigma$ and $x$ are the number density, cross section of each species and path length in  the atmosphere, respectively. Hereafter we only calculate $\rm O_2$ absorption lines because $\rm H_2O$ absorption lines largely depend on the climate variation in the transmitted atmosphere.
The $\rm O_2$ absorption line ($T_{1{\rm O_2}} (z_{min})$) made by the light passing at an altitude $z_{min}$ is expressed as;
\begin{equation}
 T_{1{\rm O_2}} (z_{min}) = e^{-\int (u_{{\rm O_2}} (z) \cdot \sigma_{{\rm O_2}} (z))dx},
\end{equation}
where $u_{{\rm O_2}} (z)$ and $\sigma_{{\rm O_2}} (z)$ are the number density and cross section of $\rm O_2$ in the atmosphere, respectively, and the summation in the exponent is taken along the ray.

The cross section has been developed with the rational polynomial of Kuntz (1997) which is revised by Ruyten (2003) used for the approximation of the Voigt profile function. 
The FWHM of Stokes function is estimated using the data of HITRAN 2012 (Rothman et al. 2013) and the FWHM of Doppler function ($\gamma_D$) is determined by the line position $\nu$, the molecular mass $m$ and the temperature $T$ as,
\begin{equation}
 \gamma_D = \nu \sqrt{\frac{2kT}{mc^2}} \approx 7.6 \cdot 10^{-8} \nu \sqrt{T \ \rm [K]}  \ \ (m = 32 \ {\rm amu}),
\end{equation}
where $k$ is Boltzmann's constant and $c$ is speed of light.

Number density is calculated from the equation of state using temperature ($T$), pressure ($P$) and  $\rm O_2$ abundance at each point along the ray. As the cross section and number density depend on $T$ and $P$, these values are calculated at each altitude assuming the vertical structure of Earth atmosphere based on the MIPAS model.
As $T$, $P$, and molecule abundance are given for every 1 km in height from 0 to 120 km from the surface in the MIPAS model, we calculate the number density every km. The number density at an arbitrary altitude is obtained by linear interpolation between these values.

We divide the ray into small segments and calculate $z$, $u$, $\sigma$ and $x$ for each segment. Then we take summation of the product of $u$, $\sigma$ and $x$ along the ray, and obtain the transmission spectrum as,
\begin{equation}
T_{1{\rm O_2}} (z_{min}) = e^{- {\displaystyle \sum^{}_{i}} u_{{\rm O_2}} (z_i) \cdot \sigma_{{\rm O_2}} (z_i) \cdot \Delta x (z_i)},
\end{equation}
where $z_i$ is the altitude of the $i$-th segment and $\Delta x (z_i)$ is the length of the $i$-th segment.

 Then, we also consider $\rm O_3$ absorption and $\rm O_2$ $\cdot$ $\rm O_2$ Collision Induced Absorption (CIA), which contribute to the spectra at broad band wavelength, around the $\rm O_2$ absorption lines (6275-6330 $\rm \AA$) and Rayleigh scattering which prevents all sun light from passing through the lowest layer of Earth’s atmosphere. We calculate these transmission spectra by combining each cross section and number density along the ray.
The cross sections of $\rm O_3$ for various temperatures (193K - 293K in steps of 10K)  are taken from the ASCII data at IUP Bremen (Gorshelev et al. 2014; Serdyuchenko et al. 2014).  According to these data, $\rm O_3$ absorption cross section has only little temperature dependence at our observational waveband (Chappuis band). Therefore, we use the cross section for 223K, which corresponds to the altitude where the number density of $\rm O_3$ reaches a maximum in the Earth's atmosphere. The number density is calculated from the equation of state using temperature ($T$), pressure ($P$) and  $\rm O_3$ abundance. The $\rm O_2$ $\cdot$ $\rm O_2$ absorption cross sections are taken from HITRAN database (Richard et al. 2012).

 For Rayleigh scattering, we utilize the number density of all particles in the atmosphere and the cross section of the Rayleigh scattering for the Earth's atmosphere from Bodhaine et al. 1999;
 \begin{equation}
 \sigma_{Rayleigh} = \frac{1.054 - 341.29 \lambda^{-2} - 0.9023 \lambda^{2}}{1+0.002706 \lambda^{-2} - 87.9686 \lambda^{2}} (\times 10^{-28} \rm cm^{2}),
\end{equation} 
where $\lambda $ is in $\rm \mu m$.

Finally, we calculate all the transmission spectra attributed to the various rays that reach the observation point on the lunar surface, and integrate them weighted by the limb darkening (quadratic law; $u1$ = 0.5524, $u2$ = 0.3637) on the solar disk. The synthetic spectrum thus obtained is shown in Figure~13.  As seen in the figure, it is deeper and wider than the observed one.

\section{Discussion}
We conducted spectroscopic observations of a eclipse using Subaru/HDS with higher resolutions in time and wavelength than previous works. As a result, we successfully detected temporal variations in individual absorption lines of $\rm O_2$ and $\rm H_2$O in Earth's transmission spectra.

The lines, however, show large difference in depth between those obtained at Point A and Point B as seen in Figure 9. Furthermore, there is apparently a discrepancy between observed and synthetic transmission spectra as seen in Figure~13. We discuss these issues in the following sections.

\subsection{Transmission spectra at Points A and B}

We alternately observed Points A and B during the lunar eclipse, and obtained a series of telluric transmission spectra for each point. As clearly seen in Figure~\ref{fig:eight}, the depth of absorption lines in the transmission spectra largely differs between the two points despite the fact that the sunlight passed through nearly the same atmospheric layers of the Earth before reaching the lunar surface (Figure~\ref{fig:eleven}). Considering the fact that the difference is not seen in the solar spectra $S$ but in the transmission spectra $T_1$, 
full-Moon and Penumbra light were scattered into Point A and increased the background making the transmission absorption lines shallower.

Actually, as shown in Figure~\ref{fig:three}, Point A is ahead of the rest  of the lunar surface toward the shadow of the Earth. Therefore, when Point A enters into the Earth shadow, the rest of the lunar surface, which is a part of the full-Moon, is still bright. On the other hand, Point B is located at the back end edge of the lunar surface, and thus the rest of the lunar surface is already faint in the shadow when Point B enters into the shadow. 

\subsection{Comparison between observed and modeled transmission spectra}
In this section, we deal with the transmission spectra obtained at both points in umbra.

\begin{figure}[htb]
\begin{center}
\vspace{0mm}
\includegraphics[keepaspectratio,width=80mm] {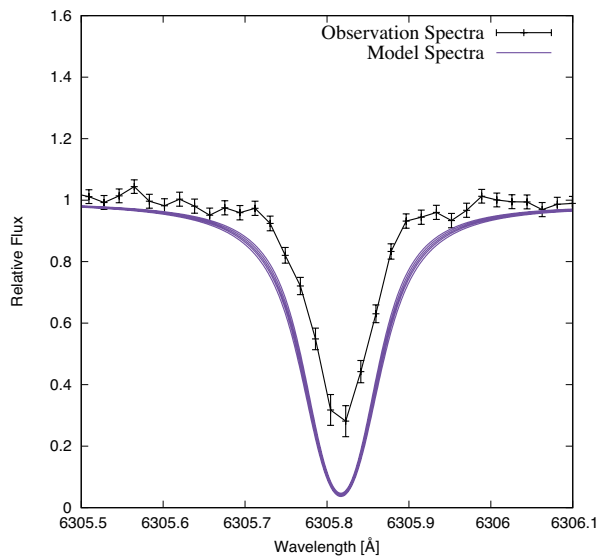}
\end{center}
\caption{Comparison between observational and modeled $\rm O_2$ absorption line for 14:47 UT when whole lunar disk moved into the Earth's shadow and the effect of the scattered light is the lowest. The thickness of these line shows the uncertainty of model spectrum.}
\label{fig:thirteen}
\end{figure}

Figure~\ref{fig:thirteen} shows a comparison between the observed and the synthetic transmission spectrum of $\rm O_2$ (section 4.2). As seen in the figure, the synthetic absorption line is deeper and wider than observed one. Considering that the observed spectrum may be affected by background scattered light, we first adjusted the line depth of the synthetic spectrum by adding a constant to it and renormalizing it to the continuum level. However, the width of the absorption line still shows discrepancy between the observed and the modeled spectrum. We discuss possible causes of the discrepancy in more details below.

The first one is uncertainty in the adopted vertical structure of the Earth's atmosphere. We used that of mid-latitude of MIPAS model since the lowest altitude at which the sunlight passed is from the center of Southern Indian ocean to the Northern coast of Antarctica. However, the actual region where the sunlight passed through is much wider (see Figure \ref{fig:eleven}).
In order to evaluate how the vertical structure affects the resultant transmission spectrum, we tested five different thermal structure models; polar-winter, polar-summer, and equatorial model of MIPAS, and mid-latitude summer and subarctic-summer model of FASCODE/ICRCCM (Clough et al. 1992). As a result, the vertical structures cannot account for the difference between the modeled and observed spectrum.

The second possibility is uncertainty in the model spectrum. 
HITRAN has the uncertainty of intensity, air-broadened half-width, self-broadened half-width, and the coefficient for the temperature-dependence of the air-broadened half-width.
In our observational wavelength, the uncertainty range of intensity is 5-10\% and that of air-broadened half-width, self-broadened half-width, and the coefficient for the temperature-dependence of the air-broadened half-width is 10-20\% or over 20\%.
The uncertainty in final model spectra caused by the factors described above was shown by Figure~13.  This shows that we cannot explain the difference between model and observation by the uncertainty in HITRAN database.
We omit the absorption lines which has the uncertainty range of over 20\% in HITRAN database from following discussion since we cannot estimate uncertainty in final model spectra.

\begin{figure}[htb]
\begin{center}
\vspace{0mm}
 \includegraphics[keepaspectratio,width=80mm] {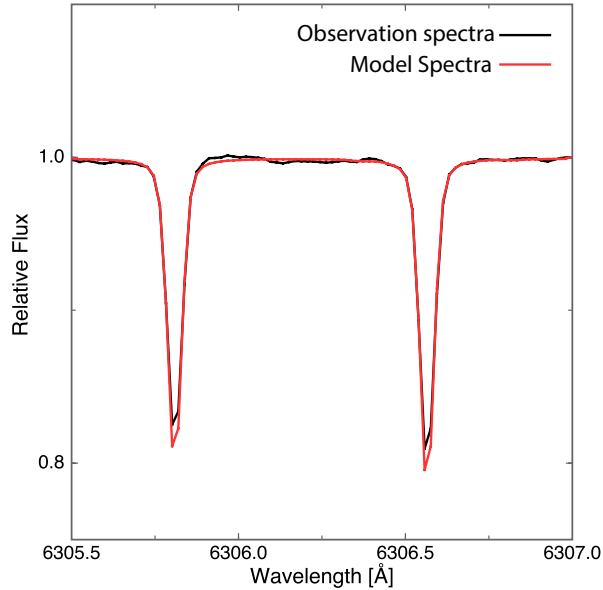}
\end{center}
\caption{The modeled (red) and observed (black) $\rm O_2$ absorption lines of the Earth atmosphere ($T_2$). The synthetic lines are only 1.1 deeper than observed ones.}
\label{fig:fifteen}
\end{figure}

Furthermore, in order to validate our calculation of the model spectrum, we tried to reproduce $T_2$ spectra, telluric spectra above the telescope, which are extracted from spectra of rapid rotators (section 3.1.1.). As a result, the modeled $T_2$ spectra are well matched with the observed ones (Figure~\ref{fig:fifteen}), and thus the uncertainty in the model spectra cannot account for the discrepancy, either.

Finally, we consider effect of clouds on transmission spectra. If clouds exist in the atmosphere, the sunlight may be blocked by the clouds and cannot reach the lunar surface. Since clouds can exist up to $\sim $13 km in the Earth's atmosphere, we may have not seen the sunlight passing the lower atmosphere, i.e. denser and higher-pressure layers. To examine this effect, we created theoretical transmission spectra that only pass through the atmosphere above some critical elevation,below which no sunlight transmits.

As seen in Figure 15, the model spectrum with the critical elevation of 9-10 km well reproduce the observed one. 
In this figure, we show only one absorption line but we obtain the same results from all observed absorption lines and all observing times. 
In addition, the model spectra with critical elevation lower than 4-5 km are identical to the original one without considering any critical elevations ( purple dotted line in Figure 15 ). This indicated that the Rayleigh scattering can prevent the sunlight from probing the lowest layer ( $<$ 4-5 km ) of Earth's atmosphere regardless of the existence of clouds. 
The result is apparently inconsistent with the previous works (Vidal-Madjar et al. 2010 and Arnold et al. 2014) showing that the Rayleigh scattering prevent the sunlight from probing the lower layer of the atmosphere of about $<$ 20 km/s at 6300 $\rm \AA$.
The inconsistency probably comes from the fact that we observed different altitude of Earth's atmosphere from the previous observations. 
As these previous works observed the limb of the Earth shadow, they can observe the upper atmosphere and most of the sunlight come from altitude higher than $\sim $20 km where extinction by Rayleigh scattering is smaller. On the other hand, as we observed the center of the Earth shadow, we can only observe the sunlight passed through the atmosphere under about 20 km. Although the amount of sunlight that pass through the atmosphere below 20 km is lower than that through above 20 km due to Rayleigh scattering, it is still detectable as shown by our observations and we can reveal detailed structure of the lower atmosphere by only probing this part.
We also prove the existence of $\rm O_2 \cdot O_2$ and $\rm O_3$ spectra have little influence with $\rm O_2$ absorption lines.

To verify our results, we examine the cloud top pressure in the sunlight-transmitted area (from the center of Southern Indian ocean to the Northern coast of Antractica) during observation time using Moderate Resolution Imaging Spectroradiometer (MODIS) data acquired from the Level-1 and Atmosphere Archive \& Distribution System (LAADS) Distributed Active Archive Center (DAAC)\footnote[4]{https://ladsweb.modaps.eosdis.nasa.gov/search/}.
These data indicate that the cloud top pressure is about 200-450 hPa (i.e. cloud top height was about 6-11km), which is consistent with our results. 

\begin{figure}[htb]
\begin{center}
 
 \includegraphics[keepaspectratio,width=7.5cm] {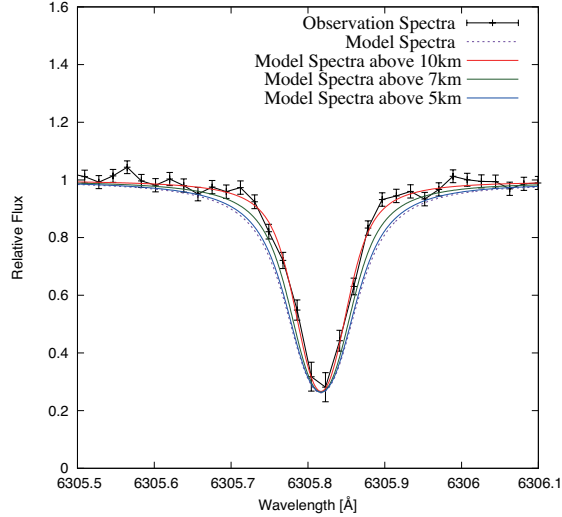}
\end{center}
\caption{The $\rm O_2$ absorption line of observation and models. This observation line (black) was observed at 14:47 UT when whole lunar disk moved into the Earth's shadow and the effect of the scattered light is the lowest, and these model lines are developed and corrected so that the depth fits the observation's, assuming that the sunlight only can transmit the atmosphere above 5 (blue), 7 (green) and 10 (red) km. }
\label{fig:seventeen}
\end{figure}

\section{Summary}

Using Subaru HDS on UT 2011 December 10, we obtained high resolution transmission spectra of Earth by fixed-point observation at two points (Point A, Point B) on lunar surface. We clearly detected temporal variations in $\rm O_2$ and $\rm H_2O$ absorption lines of the transmission spectra. In order to investigate the cause of the temporal variations, the minimum altitude ($z_{min}$), at which the lights transmitted in the Earth's atmosphere, was calculated by taking account of the atmospheric refraction at each observation time. We found that the sunlights are transmitted on the lower altitude range with time at both observation points. We also found that the spectra of Point A, which show much shallower absorption lines than those of Point B, are probably affected by the scattered light from other bright parts of the lunar surface.

Furthermore, we made a synthetic transmission spectrum of the Earth with Voigt function by taking account of the limb-darkening, refraction and dependence of the spectrum on atmospheric altitude at which the sunlight passes. The model spectra turned out to show wider absorption lines than those of the observed ones even after the correction of the line depth that may be suffered from the scattered light. We found that the width is matched with the observed one if we assume that the sunlight only transmits above 9-10 km of the Earth's atmosphere. The result suggests the existence of clouds below 9-10 km that may block the sunlight.

Our results will be useful for the investigation of atmospheric structure of an Earth-like exoplanet when the time variation of its transmission spectra can be obtained in the future.

\begin{ack}
We thank the referee Dr. Luc Arnold for many valuable comments that greatly improved this paper.
We thank the support of our Subaru/HDS observation by Akito Tajitsu, Naruhisa Takato, Daigo Tomono  and Ruka Misawa.  We also acknowledge the people who supported the proposal of observation and the people who discussed this study, especially Masato Katayama, who adviced how to obtain the longitude of the lowest point  in Earth's atmosphere. The part  of our data analysis was carried out on common use data analysis computer system at the Astronomy Data Center, ADC, of the National Astronomical Observatory of Japan. This work was supported by Japan Society for Promotion of Science (JSPS) KAKENHI Grant Number JP16K17660, JP18H01265 and 15H02063.
Y.K. is supported by the Grant-in-Aid for JSPS Fellow (JSPS KAKENHI No.15J08463) and Leading Graduate Course for Frontiers of Mathematical Sciences and Physics. The cloud top pressure above limb Earth region of interest during observation time is based on Terra and Aqua MODIS Cloud Subset 5-Min L2 Swath 5km and 10km data. These data were acquired from the Level-1 and Atmosphere Archive \& Distribution System (LAADS) Distributed Active Archive Center (DAAC), located in the Goddard Space Flight Center in Greenbelt, Maryland (https://ladsweb.nascom.nasa.gov/). We acknowledge LAADS Web search.
We acknowledge the very significant cultural role and reverence that the summit of Mauna Kea has always had within the indigenous people in Hawai'i. 
\end{ack}

\appendix
\section{Terrestrial coordinates of the sunlight-transmitting regions in Earth's atmosphere}

\subsection{Latitude}

\begin {figure} [htb]
\begin{center}
\includegraphics[width = 8cm] {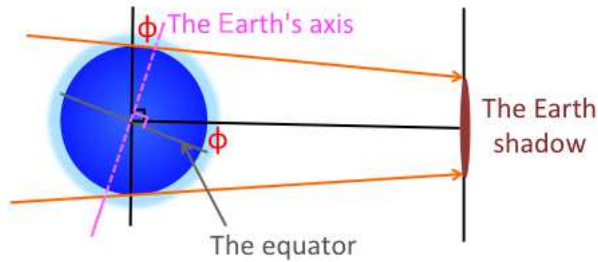}
\end{center}
\caption{A schematic view showing a relation between the axis of the Earth and the RA of the shadow of the Earth. }
\label{fig:eighteen}
\end{figure}

\begin {figure} [htbp]
\begin{center}
\includegraphics[width = 8cm] {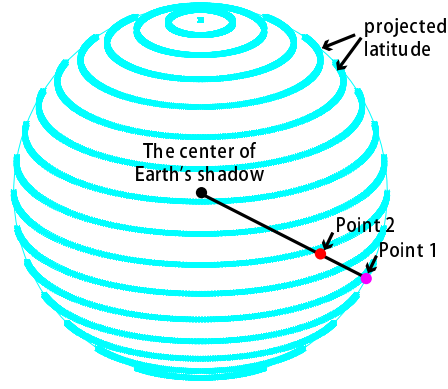}
\end{center}
\caption{A schematic view showing the projected latitude on the shadow of the Earth.  The light blue lines are the projected latitude on the shadow of the Earth, the black point is the center of shadow of the Earth, and the red and pink points are the Point A (Point 2) at 13:21 UT set on the lunar surface and the point corresponding to that the sunlight passes at the terminator of the Earth, respectively.}
\label{fig:nineteen}
\end{figure}

Earth's spin axis is inclined at about 23.4 degrees from the perpendicular to the ecliptic (Figure~\ref{fig:eighteen}). Therefore, latitudes on the Earth are projected onto the shadow of the Earth as shown in Figure~\ref{fig:nineteen}. The sunlight that passes through the Earth's atmosphere at minimum lowest altitude (point Z; Figure~\ref{fig:eleven}) on the terminator (Point 1 on Figure~\ref{fig:nineteen}) finally reaches a point (Point 2 on Figure~\ref{fig:nineteen}) on the line connecting the center of the Earth's shadow and Point 1. Thus we can know the latitude on the Earth where the sunlight passes through at the lowest altitude by reading the projected latitude of Point 1 on the Earth shadow.

\vspace{-3mm}

\vspace{-3mm}

\subsection{Longitude}

The longitude of the Point Z (Figure~12) at the Earth's terminator also changes with time because of the Earth's spin. Here we try to know the longitude of the point $\rm Z_{g}$, which is on the earth surface just beneath the Point Z, at a given point in time.

The altitude $h$ and azimuth $A$ of a celestial object at a time $t$ is expressed with the following three equations as
\begin{equation}
 \cos h \cdot \sin A = - \cos \delta \cdot \sin H,
\end{equation}
\begin{equation}
 \cos h \cdot \cos A = \cos \phi \cdot \sin \delta - \sin \phi \cdot \cos \delta \cdot \cos H,
\end{equation}
\begin{equation}
 \sin h = \sin \phi \cdot \sin \delta + \cos \phi \cdot \cos \delta \cdot \cos H,
\end{equation}
 where $\delta$ and $H$ are the declination and hour angle of the object, and here $\phi$ is the latitude of the point $\rm Z_{g}$. $H$ is expressed as,
\begin{equation}
 H = \theta_0 + t \times 1.0027379 + \lambda - \alpha,
\end{equation}
where $\lambda$ is the longitude of the point $\rm Z_{g}$, $\theta_{0}$ is Greenwich sideral time when Universal time is 0$^h$, t is observation time in UT, and $\alpha$ is the right ascension of the object. Here we regard the Point 2 in Figure~\ref{fig:nineteen} as the celestial object, which is a point on the Moon that the sunlight passing through Point $\rm Z$ reaches. The $\alpha$ and $\delta$ of the Point 2 are given by JPL's HORIZONS system\footnote[5]{https://ssd.jpl.nasa.gov/horizons.cgi} at the time $t$, and $\theta_0$ is also given by 'chronological Scientific Tables'. The $\phi$ is given by the method described above. Therefore, if we know h and A for the Point 2, we can know H and thus $\lambda$.

The $h$ is derived in the following way. The refraction angle at Point Z is the largest because Point Z is at the minimum of the lowest altitude from the surface. The ray passing through Point Z comes from the bottom of the Sun since the ray enters into the Earth's atmosphere with the largest incident angle. Therefore, at Point Z on the terminator, we see the point 2 on Moon and the bottom of the Sun in the exact opposite direction in a horizontal fashion. In addition, since the effect of atmospheric refraction makes the elevation of the point 2 seen from the point $\rm Z_{g}$  higher, we should subtract the refraction angle ($\Phi$) from the parallax ($p$) of the point 2 to correct $h$. Assuming that Point Z is $z$ km high from the earth surface, we obtain the $h$ as,
\begin{equation}
h = p - \Phi = \frac{z}{38400} - \Phi \ \ \ \ 
 (\ p = \frac{z_{Point \ Z}}{d_{EM}} \approx \frac{z}{38400} \; [{\rm rad}] \ ),
\end{equation}
where $d_{EM}$ is the distance between the Earth and the Moon. 

Next $A$ and $H$ are calculated as follows. We rewrite the equations (12)-(14) as,
\begin{equation}
\sin A = C \sin H,
\end{equation}
\begin{equation}
\cos H = F,
\end{equation}
\begin{equation}
\cos A = D \cos H + E = D \cdot F + E,
\end{equation}
where $C = - \cos \delta/\cos h$, $D = -(\sin \phi \cdot \cos \delta)/\cos h$, $E = (\cos \phi \cdot \sin \phi)/\cos h$, $F = (\sin h - \sin \phi \cdot \sin \delta)/(\cos \phi \cdot \cos \delta)$. Now all of the $C$, $D$, $E$, and $F$ are functions of the known variables. We here assume that $A$ takes the value between 0 and $\pi$ because the Moon is seen in the east from the terminator at the beginning of the lunar eclipse. 
It should be noted that the value of $H$ depends on the signs of $\sin H$ and $\cos H$ in the following way,
\begin{center}
 \begin{tabular}{cc}
i) $\cos H > 0$, $\sin H > 0$ & ii) $\cos H < 0$, $\sin H > 0$ \\
${\rm 0^\circ < H < 90^\circ}$W & ${\rm 90^\circ}$W ${\rm < H < 180^\circ}$W \\
 & \\
iii) $\cos H > 0$, $\sin H < 0$ & iv) $\cos H < 0$, $\sin H < 0$\\
${\rm 0^\circ < H < 90^\circ}$E & ${\rm 90^\circ E < H < 180^\circ}$E\\
   \end{tabular}
\end{center}   
We finally obtain $\lambda $, the longitude of the point $\rm Z_g$, from the equation (10).

Table~\ref{table:third} lists thus determined latitude and longitude of the point $\rm Z_g$, just beneath the Point Z on the earth, for Point A and B. The results show that the Point Z is above the center of Indian Ocean at Southern Hemisphere during our observations.

\end{document}